\documentclass[acmsmall, nonacm]{acmart}

\AtBeginDocument{%
  \providecommand\BibTeX{{%
    \normalfont B\kern-0.5em{\scshape i\kern-0.25em b}\kern-0.8em\TeX}}}

\setcopyright{none}
\copyrightyear{2022}
\acmYear{2022}
\acmDOI{}

\acmConference{}
\acmPrice{}
\acmISBN{}

\usepackage{xcolor}
\usepackage{booktabs}
\usepackage{graphicx}
\usepackage{graphics}
\usepackage{longtable}
\usepackage{makecell}
\usepackage{lscape}
\usepackage[normalem]{ulem}
\useunder{\uline}{\ul}{}
\begin{document}

\title[``You Just Assume It Is In There, I Guess'': UK Families' Application And Knowledge Of Smart Home Cyber Security]{``You Just Assume It Is In There, I Guess'': Understanding UK Families' Application And Knowledge Of Smart Home Cyber Security}
\thanks{This is the author version of the article: Turner S., Pattnaik N., Nurse J.R.C, Li S. (2022) \textit{``You Just Assume It Is In There, I Guess'': UK Families' Application And Knowledge Of Smart Home Cyber Security.} Accepted for publication as part of ACM CSCW 2022; publication details forthcoming.}

\author{Sarah Turner}

\email{slt41@kent.ac.uk}
\orcid{0000-0003-1246-1528}
\author{Nandita Pattnaik}
\email{np407@kent.ac.uk}
\author{Jason R.C. Nurse}
\email{j.r.c.nurse@kent.ac.uk}
\orcid{0000-0003-4118-1680}
\author{Shujun Li}
\email{s.j.li@kent.ac.uk}
\orcid{0000-0001-5628-7328}
\affiliation{%
  \institution{Institute of Cyber Security for Society (iCSS) \& School of Computing, University of Kent}
  \streetaddress{University of Kent}
  \city{Canterbury}
  \state{Kent}
  \country{UK}
  }

\renewcommand{\shortauthors}{Turner et al.}

\begin{abstract}
The Internet of Things (IoT) is increasingly present in many family homes, yet it is unclear precisely how well families understand the cyber security threats and risks of using such devices, and how possible it is for them to educate themselves on these topics. Using a survey of 553 parents and interviews with 25 families in the UK, we find that families do not consider home IoT devices to be significantly different in terms of threats than more traditional home computers, and believe the major risks to be largely mitigated through consumer protection regulation. As a result, parents focus on teaching being careful with devices to prolong device life use, exposing their families to additional security risks and modeling incorrect security behaviors to their children. This is a risk for the present and also one for the future, as children are not taught about the IoT, and appropriate cyber security management of such devices, at school. We go on to suggest that steps must be taken by manufacturers and governments or appropriate trusted institutions to improve the cyber security knowledge and behaviors of both adults and children in relation to the use of home IoT devices.
\end{abstract}

\begin{CCSXML}
<ccs2012>
   <concept>
       <concept_id>10002978.10003029.10011703</concept_id>
       <concept_desc>Security and privacy~Usability in security and privacy</concept_desc>
       <concept_significance>500</concept_significance>
       </concept>
   <concept>
       <concept_id>10002978.10003029.10003032</concept_id>
       <concept_desc>Security and privacy~Social aspects of security and privacy</concept_desc>
       <concept_significance>500</concept_significance>
       </concept>
   <concept>
       <concept_id>10003120.10003130.10011762</concept_id>
       <concept_desc>Human-centered computing~Empirical studies in collaborative and social computing</concept_desc>
       <concept_significance>500</concept_significance>
       </concept>
 </ccs2012>
\end{CCSXML}

\ccsdesc[500]{Security and privacy~Usability in security and privacy}
\ccsdesc[500]{Security and privacy~Social aspects of security and privacy}
\ccsdesc[500]{Human-centered computing~Empirical studies in collaborative and social computing}

\keywords{Internet of Things, smart home, cyber security, families, parents, children, consumer protection}

\maketitle

\section{Introduction}

A 2021 report into smart home product ownership in the UK found that 76\% of respondents owned at least one such device \citep{techuk_state_2021}. Although the picture of ownership is a complex one, with only smart TVs, smart speakers and fitness devices being reported in this survey as having ownership levels higher than 20\% of the respondent population, the increase of such devices in communal areas of the home cannot be denied. This means that all members of a household, whether adults or children, comfortable with technology or not, may be expected to live around Internet-connected devices.

Recently, research has focused upon the various usability and privacy issues of the multi-use of home IoT devices \citep{geeng_who_2019, yao_privacy_2019, bernd_bystanders_2020, kraemer_exploring_2019, chalhoub_it_2021}.\footnote{``Home IoT devices'' will be considered to be those explicitly defined in the Code of Practice for Consumer IoT Security published by the UK's Department for Digital, Culture, Media and Sport (DCMS) \citep{dcms_code_2018}.} The privacy of all people living within a home is of key importance when considering home IoT devices, and keeping data that is intended to be private out of the public domain is fundamental. The security issues arising from the adoption of home IoT devices extend beyond this, however, and the relative novelty and heterogeneity of such devices, coupled with their global production, also mean that they have, to date, evaded meaningful regulation or legislation to mandate minimum security requirements. \citet{blythe_systematic_2019} suggest that a lack of policy and technical intervention may facilitate a ``crime harvest'', where crimes targeted at the individual (burglary, sex crimes) or a societal level (such as political subjugation) will become more common; \citet{hodges_cyber_2021} has shown that homes with smart technology at present are somewhat more likely to suffer crimes such as burglary.

Home IoT device threats are much more pervasive than those of home computers, in that they can allow for enormous data collection about the users, run essential elements of a home, be linked to credit card and other sensitive information, and be leveraged for cybercrime. In a multi-user setting, device owners should be considering a range of threat actors, from malicious adversaries \cite{paul_ring_2019}, to misuse by the manufacturer \cite{statt_amazon_2018}, to those inside the house, who could use devices for control of others \citep{lopez_internet_2019}, or simply inadvertently making purchases or breaking a device. However, users' mental models of home IoT devices have been shown to be inaccurate \citep{luger_like_2016}, and this lack of understanding, coupled with a lack of interfaces, can make it harder to understand how the devices should be secured. Devices may not obviously provide users with means to implement appropriate security on the device, possibly requiring home network security to be increased for adequate protection. Home IoT devices that fail, or manufacturers that cease to exist, may cause fundamental parts of the home to stop working.

This paper extends out the work done on the privacy of multi-use devices by considering what it means for families to manage the cyber security aspects of such device use, and not solely the privacy concerns. It looks to understand how families in the UK consider and implement home IoT device cyber security, how they learn about and discuss it as a group with specific responsibilities around keeping each other safe and passing on knowledge. Using a survey of 553 individuals and interviews with 25 families,\footnote{Where a family group comprises of all the parents and/or legal guardians and school-aged children in a household (along with any other household members, although it is the parent/child interaction with devices that is of primary focus).} we investigated the concerns that families in the UK have about managing the cyber security of home IoT devices, and how and when they educate themselves about cyber security issues. Prior research in this area has tended to use either smaller interview or focus groups \citep{yao_privacy_2019, bernd_bystanders_2020, sun_child_2021}, or a survey of individuals \citep{cannizzaro_trust_2020}, or be based in the USA \citep{apthorpe_you_2020, geeng_who_2019}. By combining both a survey and interview of UK families, we are able to understand a wide range of parents' experiences of home IoT device use, along with their cyber security understanding for these devices. When combined with the qualitative detail of the interviews, we find that that families are well versed in controlling and discussing cyber security aspects that apply to being online, but do not have similar strategies in place to manage the different threats of home IoT use, assuming a certain level of protection simply from existing consumer protection regulation. Despite relying heavily on discussion as a means of facilitating a communal understanding of rules, neither parents nor children have sufficient prior knowledge or access to sufficient education or support to have meaningful conversations about home IoT device cyber security as a family unit, focusing instead on factors relating to cost saving and efficient use. We argue that this is a concern not only for the present day, but for the future as well, as children are not having appropriate behavior modeled to them by their parents, and make suggestions for the role of different stakeholders in managing this knowledge gap.

The rest of the paper reads as follows: an overview of related work is found in Section~\ref{sec: related_works}, methods of both survey and interviews are explained in Section~\ref{sec:methods}. Survey findings, followed by interview findings, are in Section~\ref{sec:findings}; these findings are discussed in Section~\ref{sec:discussion}. Limitations and further work are considered in Section~\ref{sec:limitations_future_work}, which are followed by concluding thoughts in Section~\ref{sec:conclusions}.

\section{Related Work}
\label{sec: related_works}

\subsection{Multi-use of IoT devices}

\citet{sun_child_2021} found, from interviewing 23 parents in the US and Canada, that parents' perception of the risk assessments to keep their children physically and digitally safe in the smart home typically moved through three phases, encompassing purchasing decisions, re-evaluation as the children start to use the device, and further adjustments as the children grow.  More specifically, \citet{garg_when_2019} explored how parents of different socio-economic groups and ethnicities had different concerns over their children's use of smart speakers, ranging from managing the children's use of the device, to behavioral and privacy considerations. \citet{geeng_who_2019} found that home IoT devices often created tensions, requiring ongoing cooperation and negotiation between members of the home, particularly between parents and children.  When interviewed, parents whose children used fitness trackers designed for children found that such trackers required significantly more parental interaction than expected because of the limited agency these devices provided to the children, despite device claims and parental hopes of increased child independence \citep{icsil_raising_2020}. \citet{apthorpe_you_2020} found, via a study involving 13 interviews and a survey of 508 individuals, that the lack of nuanced sharing solutions for home IoT devices meant that households were required to fall back to ``social resolution techniques'', such as agreeing not use particular features, in order that ongoing use remains acceptable for all involved. Agreement and discussion from all participants about the goals and requirements of a device is a fundamental part of what \citet{kraemer_exploring_2019} refer to as ``group efficacy'', a key element of acceptable multi-user device management. However, power dynamics need to be recognized as a limiting factor in group efficacy: \citet{bernd_bystanders_2020} report that nannies working within a family context did not consider themselves to have enough power in the relationship with parents --- their employers --- to ask for smart surveillance systems to be switched off. \citet{ehrenberg_the_2021} similarly found that the use of smart technologies in the home can work to create hierarchies by affecting access to, use of and interactions within physical household spaces.

\subsection{Parent-child communication around cyber security and privacy}

There is suggestion that communal decision making about how and when families should use Internet-connected devices (in particular, smartphones) can create higher levels of willing compliance \citep{hiniker_not_2016, ko_familync_2015}. This is in keeping with the argument of \citet{blum-ross_parenting_2020} that family life is, increasingly, a democratic one, that requires more complex rule-making around technology use than time or content restriction, and the recognition that children are beings with their own agency.

\citet{shin_parental_2015} and \citet{kumar_no_2017} found that parents will often put off discussing security with their children, erroneously considering it to be a future risk. \citet{wisniewski_parents_2017} showed that this lack of interaction between parents and teenage children in regards to how to manage Internet-related risks sets children up poorly to manage them, should they occur. Additionally, children tend not to raise instances where they have encountered risks to their parents, and feel that their parents disproportionately tell them what they cannot --- rather than can --- do online \citep{blackwell_managing_2016}.

\citet{nikken_guiding_2015} showed that parents do not understand enough to feel comfortable in what their children are doing, or how to manage it appropriately, or by themselves, often relying on external support, or older siblings. \citet{cranor_parents_2014} discussed how, for teenagers, parents struggle with the need to exhibit trust to foster maturation and independence --- but consider that their children need far less privacy in the digital realm than real life. This finding is echoed by \citet{ur_intruders_2014}, where parents agreed that children should have privacy when interviewed, only to deny them this privacy when subsequently setting up a security camera, as they should have nothing to hide.

\subsection{Cyber Security and Privacy Knowledge and Action}

In a study of 13 families, \citet{muir_exploratory_2020} found that families worried about a wide range of cyber security issues, but considered threats such as cyber bullying and stranger danger of more concern than financial or technical threats, in terms of the immediate danger to the family. When interviewed by \citet{tanczer_emerging_2018}, academic and industry participants with expertise in the security of IoT devices considered there to be a high potential for crime, exploitation, risk to physical safety and a loss of personal control to emanate from IoT devices.

It has long been understood that people form inaccurate security stances based upon inaccurate mental models of the digital technologies that they use \citep{wash_folk_2010, forget_or_2016}. \citet{huang_amazon_2020} showed that, in the case of smart speakers at least, incorrect mental models about how the devices use and present data to all users can cause individuals to make poorer decisions about privacy than they may believe they are. \citet{seymour_informing_2020} found that visualisations of home IoT data flows did not provoke change of device use or restriction of those flows, despite the information about data flows being interesting to review.

In conjunction with inaccurate understandings of the risks the devices may pose, the perceived convenience of IoT devices  leads individuals to exhibit the privacy paradox: despite considering themselves privacy conscious, in practice, they will behave in a more risky manner, in particular sharing a significant amount of personal information with home IoT devices and further with online services, where the perceived benefit of using such devices is worthwhile \citep{williams_privacy_2017}. \citet{zheng_user_2018} found that individuals trust manufacturers to protect their privacy, but do not take steps to verify this. Furthermore, users may turn to the Internet for advice on cyber security issues relating to home IoT devices, but often struggle to find formal guidance when online \citep{turner_when_2021}. In situations where accounts are shared, participants in \cite{watson_we_2020} suggested that they avoided discussing issues of security and privacy, despite the potential risks to everyone using the account. This accords with the performative aspect of trust that people exhibit when sharing devices with people that they are close to \citep{matthews_she_2016}. This lack of consistent behavior may cause problems for the next generation: \citet{kumar_no_2017} explored the mental models that children aged 5-11 had around privacy and security in relation to Internet use. While they found that the children were able to develop some strategies for control (such as providing false information), they relied heavily on their parents for support.

\section{Methods}
\label{sec:methods}

\begin{figure}[ht]
\centering
\includegraphics[width=\linewidth]{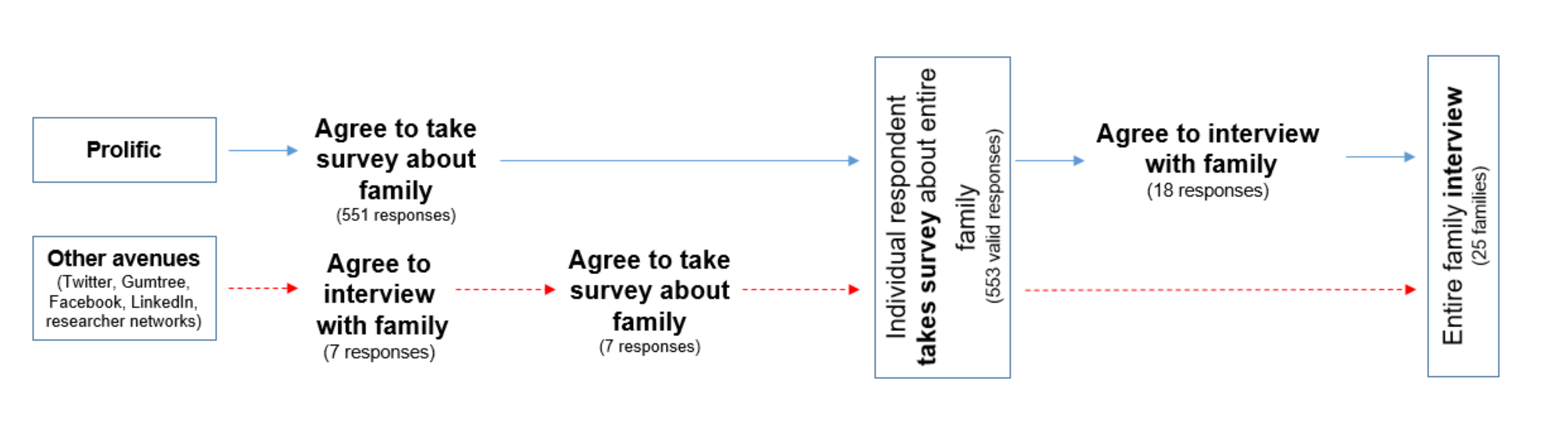}
\caption{Recruitment Process}
\label{fig:recruitment}
\end{figure}

We started our research with two main research questions: 1) how do families manage the cyber security related to their home IoT device use, and 2) how and when do families educate themselves about home IoT device security. In order to answer these questions, we undertook a survey and semi-structured interviews. For both, families had to be living in the UK, with at least one IoT device at home. Survey participants had to be a parent or legal guardian of at least one school-aged child (between 4-18); for the interview, the participants had to be a family (of at least one adult and one school-aged child). The survey was framed in order to gather a wider picture of what devices families in the UK have and what steps they take to keep those devices secure, based upon the view of a single adult family member. It additionally asked questions about the participants' experience of cyber security in the news, what their children learned at school, and what they discussed at home. The interviews followed the survey, to dig deeper into questions around the family experience of home IoT device cyber security, giving the opportunity for both parents and children to discuss. Interview participants were recruited independently, with the agreeing adult asked to take the survey prior to the family interview, or through an expression of interest after completing the survey. For a visual guide to the recruitment process, see Figure~\ref{fig:recruitment}; this is further explained in the sections that follow below. Both pieces of research received approval from the University's relevant ethics committee in July 2020.

\subsection{Survey}
\label{sec:survey_methods}

\subsubsection{Survey Design}

The survey was designed to capture a broad range of information about the type of home IoT devices found in the homes of respondents, how and who bought, used and managed them, as well as understanding the attention paid to cyber security in the home. Unlike the interviews, it was understood that only a single adult member of the family would be providing their interpretation of the family's use of such devices, giving the opportunity to ask specific questions (such as ``Does one person take ownership of managing the device?'') to understand the adult's role in family technology use. Each question gave multiple options for response, including a free text field for adding in options not covered.

The survey asked participants to list all their home devices, based upon the list found in \citep{dcms_code_2018},\footnote{This list was used as it forms the basis of devices captured by a comprehensive code of practice for ``consumer IoT devices'', set out by the UK government, and so provided a useful yardstick as to the devices available to consumers in the UK at the time. Participants were able to add in any devices they did not feel were covered in the list.} and then required the participants to answer a range of questions on the one or two devices they used most within the home. Participants were asked to focus on the most commonly used device(s) as a means of ensuring most, if not all, of the questions subsequently asked, would have been considered at some point during ownership of the particular device(s). Questions were asked about the devices that covered entire family use, from point of purchase onwards (e.g. ``Was there discussion about buying the device amongst the household members (including children) before it was bought?'', ``Does everyone in the household use the device in the same way?'', ``Do you have to help the child(ren) in the house to use the device?''). These questions were asked to provide an understanding of how the participant perceived the role of home IoT device use in the home, and where children in particular fitted in the decision making and use of such devices.

The remainder of questions about the specific devices covered the understanding of cyber security and devices, broadly based on guidance provided by the UK's National Cyber Security Centre (NCSC) \citep{ncsc_smart_2019}. These questions asked about the setup of the device (``When the device was set up, were there detailed instructions?'', and ``Did setup involve changing the device's password?''), the current use of the device (``Do you know how you would completely delete the information the device has collected?'', and ``Do you know how long for, or until when, the software on the device is supported until?'') as well as more general questions around device misuse and breaking (``Has your current device ever broken?'', and ``If the device broke tomorrow, what would be your major concerns?'') to assess cyber security risks  against other risks that the users perceived.

The final part of the survey asked for details on the participants' experience with cyber security issues more directly: ``Who or where do you turn to for support with digital technology issues?'', followed by ``Have you ever been a victim of a cybercrime of data breach?''. If they had, they were then asked if they incurred a loss as a result of this (with loss intentionally not being defined as purely financial). Finally, they were asked if they were aware of the manufacturers of devices in their home having reported a breach. In relation to cyber security and the family, they were then asked what aspects of cyber security they understood their children were taught at school, and what they discussed at home, based on the list of topics covered in the English curriculum \citep{dfe_national_2013}.\footnote{Within the United Kingdom, each of the devolved nations have their own powers to build school curricula. The curricula within Wales, Scotland and Northern Ireland cover similar ground, focusing upon online safety and learning about the role of the Internet within society in particular; however, to avoid complexity, the English Department for Education curriculum has been used in this research as the basis upon which questions are asked as it is the oldest, and is referenced as a basis for at least some of the more recent devolved nations' curricula.} These questions aimed to give a reflection of the awareness of the participant of their role as a potential victim of cybercrime, and educator about cyber security in the home.

\subsubsection{Survey participation}

The online survey was hosted on Jisc's Online Surveys tool,\footnote{\url{https://www.onlinesurveys.ac.uk/}} using the participant recruitment platform Prolific.\footnote{\url{https://www.prolific.co/}} In total, 558 adults responded to the survey during July and September 2020; 551 of these through Prolific, and 7 from the interview recruitment process. Two participants were excluded for responding solely about a laptop computer and smart phone, not a home IoT device, leaving 556 responses for further screening. There were two attention checks in the survey: four participants failed both attention checks (0.72\%); one of these participants was also interviewed. The three participants who failed the attention checks and were not interviewed were excluded from further review. The decision was taken to retain the individual who was interviewed as it was possible to see the consistency in answers between the interview and survey. This left 553 valid responses for the final analysis. Survey participants recruited through Prolific were paid pro-rata for their time, using the UK's Living Wage Foundation calculated 2019/2020 rate of £9.30 per hour\footnote{\url{https://www.livingwage.org.uk/what-real-living-wage}} (with the majority of participants taking around 15 minutes to complete it). Those survey participants recruited outside of Prolific (as part of the interview process) were paid £5 in Amazon gift vouchers for completing the survey.

\subsubsection{Demographic Information}

The participants of the survey had a mean age of 42.32 years (SD 7.51), with an age range of 24-66. 138 (24.82\%) were male, 417 (75.00\%) female, with one participant preferring not to answer this question. All participants lived in the UK at the time of taking the survey, with 473 living in England (85.07\%), 41 in Scotland (7.37\%), 21 in Wales (3.78\%) and 18 in Northern Ireland (3.24\%); although not intentionally sought, this proportion is broadly in line with the population breakdown across the countries in the UK as reported by the UK's Office of National Statistics for 2020 \citep{ons_population_2021}.
81.69\% of participants were employed either full- or part-time, or due to start a job within the next month, with 2.51\% considering themselves to be unemployed, but job-seeking. The respondents were asked to detail out the age of their children. 17.36\% of respondents (96) had a 0-5 year old child, 54.07\% (299) had a child aged 6-10, 53.52\% (296) had an 11-15 year old, and 19.89\% (110) had a 16-17 year old.

\subsection{Interview}
\label{sec:interviews}

\subsubsection{Interview Design}

The interview process was designed to build upon and flesh out the survey questions, by allowing the initial survey participants to expand upon their survey responses, and also to enable the entire family to add their views. The interview format was semi-structured, meaning that each individual could spend time focusing more on areas of concern (or awareness) for them; the full set of questions used in the interviews can be found in Appendix~\ref{appendix_interview_questions}.

The interview process was designed in such a way that the children would be asked questions by the interviewer first, where possible, without interruption from their parents; the parents were then asked to answer the questions after the children had spoken. This allowed for the parents to reflect on the children's answers, as in most cases, the parents had remained within earshot as their children were interviewed. The questions for children were modified based on whether the child was either in primary school (4-11 years) or secondary or tertiary education (12-18 years) in two ways: the expected level of knowledge was taken into consideration, based upon the national curriculum for England laid out in \citep{dfe_national_2013}, but also, the language used was modified by the interviewer based upon the age of the child. Questions focused on how each participant used the devices they chose to talk about: which device features did they like and dislike using, where they struggled to use the devices (and how they managed that). Participants were then asked, more specifically, about their understanding of what cyber security was. Children were asked to reflect on their understanding of what cyber security was, whether from school or elsewhere and what they considered the biggest risks of using the Internet, and home IoT devices were. Parents, additionally, were asked to share any situations of being a victim of a cyber crime or incident, any times where they felt that they needed to seek further information, and how they went about it.

\subsubsection{Interview Participation}

There were 25 interviews, taking place between July and September 2020; the longest interview was 1 hour 15 minutes, with the shortest being 20 minutes long. There were 33 adults interviewed in total, with 22 being female (66.67\%), 11 male (33.34\%). There were 38 children interviewed, aged from 4-16 (average age: 9.63 years, SD 3.59). Given ongoing COVID-19 social distancing restrictions at the time of interview, three video-conferencing systems were used: Microsoft Teams, Zoom and Google Meet, based on the participants' preference. Where possible, all family members were asked to attend, in particular, to enable the interviewer (the first author of the paper) to ask questions of all family members and avoid the ``driver'' of device use answering all the questions on behalf of less knowledgeable family members, and to understand the interaction the family had when discussing the technology. This was not always possible, but for all interviews at least one adult and one child from the family participated. Participants were sourced through posts on social media and message board sites (Facebook, LinkedIn, Twitter, Gumtree), through expressing interest having completed the survey, and through the first author's networks. Interview participants were remunerated with a £10 Amazon gift voucher if aged 18 or over, and a £5 Amazon gift voucher if under 18.

The participant details are in Table~\ref{tab:interview_participants}; throughout this paper, adults will be referred to through the following coding system: A (for adult), M or F (male or female), and their family's Interview Reference number, e.g., the mother in interview 2 is referred to as ``AF2'', the father in interview 3 as ``AM3''. Children are referred to by C, their family's Interview Reference number, then distinguished by their age, e.g., the two children in interview 14 are referred to as ``C14, aged 13'' and ``C14, aged 16''.

\begin{table}[t]
\caption{Information of interview participants}
\label{tab:interview_participants}
\scalebox{0.8}{
\begin{tabular}{@{}ccc@{}}
\toprule
\textbf{Interview Reference} & \textbf{Number of adults} & \textbf{Number and ages of children} \\ \midrule
1 & 1 & 1 (15)\\
2 & 2 & 2 (12,15)\\
3                            & 1                         & 1 (5)                                \\
4                            & 2                         & 1 (4)                                \\
5                            & 1                         & 1 (10)                               \\
6                            & 1                         & 1 (6)                                \\
7                            & 2                         & 3 (7, 9, 11)                         \\
8                            & 1                         & 2 (5, 6)                             \\
9                            & 2                         & 2 (4, 8)                              \\
10                           & 2                         & 3 (7, 12, 13)                        \\
11                           & 1                         & 2 (7, 14)                             \\
12                           & 1                         & 1 (13)                               \\
13                           & 1                         & 1 (12)                               \\
14                           & 2                         & 2 (13, 16)                            \\
15                           & 1                         & 1 (7)                                 \\
16                           & 2                         & 2 (7, 9)                              \\
17                           & 1                         & 1 (8)                                 \\
18                           & 1                         & 1 (10)                               \\
19                           & 1                         & 1 (12)                                \\
20                           & 2                         & 2 (6, 9)                              \\
21                           & 1                         & 1 (9)                                \\
22                           & 1                         & 1 (8)                                \\
23                           & 1                         & 2 (14, 15)                            \\
24                           & 1                         & 1 (8)                                \\
25                           & 1                         & 1 (16)                               \\ \bottomrule
\end{tabular}}
\end{table}
\subsubsection{How the children participated}

The children interviewed had a wide range of ages, and so their participation and interaction in the interview process differed. All children reacted reasonably well to the use of video-conferencing and talking to the interviewer; in two circumstances, the children asked not to be interviewed on camera, which posed no problems, as they were comfortable to talk whilst off camera. In many cases, the opening questions, asking the children to describe the home IoT devices (more details on how this was handled by children of differing ages in the paragraphs below), followed by questions asking how they used them, were enough to allay any initial reticence to participate.

The youngest children (aged 4-5) did not sit by themselves to be interviewed: being interviewed in their own homes allowed them to wander around, and in certain circumstances, answer questions by doing --- giving demonstrations of how they interacted with devices, rather than giving verbal explanations. Parents of children in this age group often re-framed questions posed by the interviewer, not only to focus the child upon answering the question but also to use more familiar terms (for example, ``how do you use the device?'' was typically modified by parents in ways such as ``how do you use Alexa?''). Parents of children in this age group supplemented answers with additional descriptions of device use and their experience of their children's level of understanding of how to use the devices. Children in this age group were able to think about, and talk through, how the devices they were familiar with might work --- or make clear to the interviewer that they did not know.

Children roughly between 6-10 would participate with a parent on- or off-screen. This provided reassurance to the child that it was safe and appropriate to answer the questions posed by the interviewer (who was, of course, a stranger to them). In particular, children in this age group would not always give full answers to questions, so parents would prompt fuller explanations from their children when they recognized this was needed. Children of this age did not seem particularly inhibited by their parents' presence, with at least one child quite happily discussing talking to strangers online, much to the shock of their parents, despite sitting right next to them. Children were able to think through how devices might work, apply concepts they had started to learn at school, and had some idea of risks that might come from the Internet, if not home IoT devices.

Children older than 10 or 11 were able to hold a conversation by themselves with the interviewer, with little to no interaction with the parents as they were interviewed. In the case of children in this age group, many commented that they had become used to talking this way as a result of COVID-19 lockdowns and home schooling in particular. Older children often also listened in to their parent(s)' responses, which sometimes promoted broader discussions between the parent(s) and the child(ren) about the question being asked.

Interviews were carried out by the first author of the paper, audio recorded and transcribed. The transcripts from the interviews were subjected to inductive thematic analysis \citep{braun_thematic_2006} by the first two co-authors of the paper. Such thematic analysis allows relative freedom for researchers to distinguish, and then refine, themes as the data is reviewed, rather than starting from a point of pre-determined coding framework. This provides a flexibility that is important when trying to ascertain significant themes that may otherwise be overlooked. The authors undertaking the coding performed the process entirely independently, met to review similarities in the generated codes, and discussed differences. After the discussion, both coders refined their code books individually, and a subsequent analysis of similarity generated a Cohen's \textit{kappa} value of 0.63, indicating a substantial level of agreement. The agreed-upon codes, with description of primary sub-codes, can be found in Appendix~\ref{appendix_code_book}.

\section{Findings}
\label{sec:findings}

\subsection{Survey}
\label{sec:survey_findings}

Participants reported owning 2,326 devices in total: the top ten devices owned overall are reported in Table~\ref{tab:all_devices_top_ten}. There are three types of devices that appear to be much more prevalent than others: smart TVs, smart speakers/home assistants and streaming devices (such as Google Chromecast devices and Amazon Fire sticks), with 1,181 of these three devices being recorded. 

Participants subsequently gave detailed information on 886 of these devices: the top eleven types of devices where participants gave detailed information reported in Table~\ref{tab:detailed_devices_top_eleven}. 

\subsubsection{Summary of findings}

This paragraph gives a high-level overview of the survey findings presented below. Devices that are usually more expensive or integral to the household (e.g., Smart TVs, refrigerators) were reported as being less new than other devices.  Participants mostly bought devices after reading online customer reviews; despite over half responding that they received detailed instructions, the majority of participants could not provide significant details about cyber security requirements, either at set up or throughout device life.  Participants' primary concern about the device breaking was around the cost of replacement, not the implications for data security.

Participants reported, on the whole, neither helping the children in the household with the device, or having concerns about children using the devices. This was especially the case for older participants, or where there were older children in the family.

Participants overwhelmingly reported getting information for cyber security information online; the majority of participants were not aware of having been the victim of a data breach or cyber crime, or if the manufacturers of any of the devices in their home had ever reported a breach. Reported learning at school and that discussed at home focused more upon online safety than cyber security steps: despite men being more likely to say they managed devices in the home, they were less likely to show knowledge of the children's school curriculum. Older participants seemed to be more likely to discuss cyber security strategies (rather than online safety) with their children.

\subsubsection{Quantitative Analysis}
 
As a result of the largely categorical nature of the variables in the survey, alongside descriptive statistics, the results were analyzed using chi-squared tests of independence to determine any associations between three of the main demographic variables and different survey variables. The demographic variables considered were: the age of the participants (based on the age ranges of 30 and below, 31-40, 41-50, 51 and above) and the age of the oldest child (based on the age ranges of 0-5, 6-10, 11-15, 16-18) and the gender of the participant responding to the survey (male, female). 

For the purposes of the chi-squared tests of independence, one record, of the one participant that answered ``prefer not to say'' in relation to their gender, was excluded as an outlier so as not to unduly influence the results. The sample size ($N$) used for calculation is different depending upon whether the question responded to was about a specific device, or about the participant and their family in general. For questions relating to the device, the sample size is the responses provided ($N = 884$); for questions relating to the participant and their family, the sample size is the number of participants ($N = 552$). The alpha value used throughout is $< 0.05$.

We had the following hypotheses:

\textbf{The null hypothesis:} There is no correlation between the particular demographic groups and how parents behave, in relation to their children and their use of IoT devices, portrayed through the survey question's variables.

\textbf{The alternative hypothesis:} There is a correlation between the specific demographic variables and the compared survey variables.

Those results that were statistically significant, and where, therefore, we could reject the null hypothesis, are detailed below.

\subsubsection{Responses about devices}
\paragraph{\textbf{Age of devices}}

Although there was a relatively even spread in the age of devices (0-6m: 18.40\%, 6-12m: 24.94\%, 1-2 years: 27.43\%, 2+ years: 29.23\%), it was notable that 139 of the Smart TVs (40.17\% of the 346 smart TVs considered in the detailed questions) were recorded as being over 2 years old. Although with much smaller reported numbers, 100\% of connected refrigerators/freezers (3 reported) and 55.56\% of connected washing machines and tumble dryers were bought two years ago or earlier (5 of 9 reported). These larger devices, which cost more and are considered to be a standard fixture of a family home, are not only less new than many of the other devices in the survey, but were reported to be bought in spite of the smart/connected features frequently. 2 of the 3 refrigerator/freezers were not bought for the smart features, along with 4 of the connected washing machine/tumble dryers, and 66 (19.08\%) of the smart TVs.

\begin{table}[ht]
\caption{Top ten reported device types}
\label{tab:all_devices_top_ten}
\resizebox{\columnwidth}{!}{%
\begin{tabular}{@{}ccc@{}}
\toprule
\multicolumn{1}{c}{Device type} & Number owned & Percentage of all devices \\ \midrule
Smart TV & 437 & 18.57\%\\
Smart speaker + home assistant & 414 & 17.55\%\\
Device connecting smartphone to TV (e.g., Chromecast) & 330          & 14.02\%                   \\
Smart meter                                          & 134          & 5.69\%                    \\
Smart printer                                        & 130          & 5.52\%                    \\
Connected children's toy                            & 107          & 4.55\%                    \\
Smart lighting                                       & 96           & 4.08\%                    \\
Connected smoke detector                             & 89           & 3.78\%                    \\
Connected thermostat                                 & 84           & 3.57\%                    \\
Connected doorbell                                   & 82           & 3.48\%                    \\ \bottomrule
\end{tabular}
}
\end{table}

\begin{table}[ht]
\caption{Top eleven devices: detailed participant response}
\label{tab:detailed_devices_top_eleven}

\resizebox{\columnwidth}{!}{%
\begin{tabular}{@{}ccc@{}}
\toprule
Device type & \makecell{ Number in depth \\ questions answered about} & Percentage \\
 \midrule
Smart TV                                                           & 346                                                                                                     & 39.05\%                        \\
Smart speaker + home assistant                                     & 207                                                                                                     & 23.36\%                        \\
Device to connect smartphone to TV (e.g., Chromecast, Fire Stick) & 106                                                                                                     & 11.96\%                        \\
Connected doorbell                                                 & 39                                                                                                      & 4.40\%                         \\
Connected children's toy                                           & 33                                                                                                      & 3.72\%                         \\
Smart meter                                                        & 29                                                                                                      & 3.27\%                         \\
Connected thermostat                                               & 18                                                                                                      & 2.03\%                         \\
Smart printer                                                      & 18                                                                                                      & 2.03\%                         \\
Smart camera                                                       & 16                                                                                                      & 1.81\%                         \\
Connected smoke detector                                           & 13                                                                                                      & 1.47\% 
\\ Smart lighting                                          & 13                                                                                                      & 1.47\%                         \\\bottomrule
\end{tabular}
}
\end{table}

\paragraph{\textbf{How devices were researched prior to purchase}}

Respondents gave some insights into how they found reviews for the devices they were answering for. Participants either did not, or did not remember, researching 226 (25.51\%) of the devices. Across the remaining devices, 572 (64.56\%) devices were bought after reading ``customer reviews on the Internet''; this option far outweighed more typically trustworthy means of ascertaining reviews, with only 116 devices (54 of which were smart TVs) being bought after the participant read a review from a consumer protection body such as Which?, the UK consumer protection body.\footnote{\url{https://www.which.co.uk/}} Much more important seems to have been the cost of devices relative to alternative products (this being a consideration for 316 of the devices), as well as what the device looked like (a consideration for 226 of the devices), and whether the device was compatible with other devices in the home already (200 devices).

\paragraph{\textbf{Who managed the devices?}}
Devices reported on in the survey were typically managed by one individual, whether adult or child (633 devices; 71.44\%). The decision for one person to manage these devices was reported as being intentional in 59.87\% of the cases (for 379 of those 633 devices). There was a relatively equal split between whether people intentionally decided how their device would be managed, with participants either confirming that there was no active decision making process for management (or that they could not remember this decision being made) for 428 of the devices (49.42\%). Despite this, there were only 6 reported devices where the status quo had caused issues, ranging from disagreement over device use, to forgetting passwords (0.68\%). There was a statistically significant relationship between gender and reported management of the device\footnote{With the options for management of the device being managed by: ``me'' (the participant), ``another adult'', ``a child'', ``no one person in the household'' or ``a landlord''.} ($\chi^2(4, N = 884) = 57.21$, $p< .001$), with the data showing that men were more likely to say they had sole ownership of managing the device (62.8\% for men against 39.3\% for women).

\paragraph{\textbf{Security at the point of device setup}}
The survey asked participants if they had received ``detailed instructions'' about how to set up their device. They responded that 536 of the 886 devices (60.50\%) did have, with physically printed instructions included with 365 of the devices; 151 devices had prompts to visit a website, the remainder had a combination of both options. In one case, the instructions were also given by the retailer. The phrase ``detailed instructions'' was left deliberately vague for participants to interpret as they found appropriate. The survey showed that across all participants an overwhelming majority of users did not know when their devices were supported until --- this was not known for 843 (95.15\%). Having ``detailed instructions'' made little difference to this figure: of the 536 with such instructions, participants reported not knowing the supported life of 505 (94.22\%) of them.

The majority of users did not remember set up involving a password change: 495 devices (55.87\%) did not require a password change, and respondents did not remember for 196 of them (22.12\%). Other activities that are linked to the security and privacy of the device are also poorly understood by users: respondents did not know how to delete personal data from 685 devices (77.31\%). There was a statistically significant relationship between gender and whether the participant knew how to delete personal data from the device or not ($\chi^2(1, N = 884) = 7.97$, $p < .005$); men were more likely to report understanding how to delete data from the device (29.6\% for men; 20.4\% of women). When asked to give their concerns, should the device break tomorrow, participants were overwhelmingly concerned about replacing the device (67.16\% of all devices), not over whether the data they had provided would become inaccessible to them (12.87\% of all devices), or subsequently deleteable (0.23\%, just 2, devices).

\subsubsection{Family Use}

Respondents were asked about the type of help that they provided their children to use for each device they answered questions about. For 687 of the devices (77.54\%), respondents suggested that they felt no need to give help to their children. When they did give help, parents reported explaining how to interact with the device the most (93 devices, 10.50\%); in the case of 83 devices (9.37\%), respondents suggested that they managed their children's usage to avoid physical damage, or damage to the service the device provided. Respondents felt that they had no concerns about their children using the devices for 572 (64.56\%) of those devices surveyed, and indeed, children were reported as using 682 of the devices (76.98\%) by themselves, with very few reported issues.

\paragraph{\textbf{Older participants provide less help and have fewer concerns}}The age of participants was shown to be statistically significant in terms of both helping children with devices\footnote{With the options for``who helps the children in the home'' being ``all/most of the adults in the home'', ``another adult in the home'', ``no one in the home'', ``just me in the home'', or ``I don't know'' (where `me' and `I' refer to the participant).} ($\chi^2(12, N = 884) = 67.18$, $p < .001$), and being concerned that children in the home may damage the device\footnote{With the options being ``I have concerns'', ``I do not have concerns'', or ``the children do not use this device''.} ($\chi^2(6, N = 884) = 31.21$, $p < .001$). In particular, the amount of help given to children seems to decrease as the age of the parent increases, and the same is true of the concern around breaking the device. For further detail, see Table~\ref{tab:age_participant}. 

\begin{table}[ht]
\caption{Age ranges of adult participants and interactions with their child(ren) (percentage based on all responses to the question by age range)}
\centering
\resizebox{\textwidth}{!}{%
\begin{tabular}{@{}ccccccccc@{}}
\toprule
\multicolumn{1}{c}{\begin{tabular}[c]{@{}c@{}}Age range \\ of participant\end{tabular}} &
  \multicolumn{1}{c}{\begin{tabular}[c]{@{}c@{}}Children do not \\ get help with devices\end{tabular}} &
  \multicolumn{1}{c}{Percentage} &
  \multicolumn{1}{c}{\begin{tabular}[c]{@{}c@{}}I am not concerned about \\ children breaking the device\end{tabular}} &
  \multicolumn{1}{c}{Percentage} &
  \multicolumn{1}{c}{\begin{tabular}[c]{@{}c@{}}Children learn about cyber \\ security topics that could be \\ applied to devices at school\end{tabular}} &
  \multicolumn{1}{c}{Percentage} &
  \multicolumn{1}{c}{\begin{tabular}[c]{@{}c@{}}I talk to children about cyber \\ security topics that could be \\ applied to devices at home\end{tabular}} &
  \multicolumn{1}{c}{Percentage} \\ \midrule
30 and under & 28  & 52.8\% & 26  & 49.1\% & 5   & 20.8\% & 14  & 58.3\% \\
31-40    & 264 & 75.0\% & 199 & 58.5\% & 68  & 30.2\% & 125 & 55.6\% \\
41-50    & 302 & 79.3\% & 256 & 67.2\% & 113 & 47.1\% & 165 & 68.8\% \\
51 and over  & 92  & 92.0\% & 82  & 82.0\% & 42  & 65.6\% & 49  & 76.6\% \\ \bottomrule
\end{tabular}%
}

\label{tab:age_participant}
\end{table}

\paragraph{\textbf{The presence of older children makes for less help or concern for all children}}Similarly, the age of the oldest child in the family seems to make a difference to how the participant reported the help they provided or concerns they had for all of their children. When chi-square tests of association were applied to the eldest child a participant reported as having, there were statistically significant outcomes as to the generally reported levels of help given to children ($\chi^2(12, N = 884) = 65.37$, $p < .001$), and concerns about breaking or damaging devices ($\chi^2(6, N = 884) = 36.36$, $p < .010$), as well as whether the participant believed their children were competent with the device or not ($\chi^2(3, N = 884) = 16.77$, $p < .001$). Survey participants with children in the oldest age range (16-17) were less likely, based upon the responses in the survey, to report giving help to any children in the family, or having any concerns about the children damaging the device. Interestingly, only when the oldest child was 5 or under was there a clear difference in the likelihood that the participants considered the children competent with the device. See Table~\ref{tab:age_oldest_child} for more detail.

\begin{table}[ht]
\caption{Age of oldest child and perceived ability (percentage based on all responses to the question by age range of oldest child)}
\centering
\resizebox{\textwidth}{!}{%
\begin{tabular}{@{}ccccccc@{}}
\toprule
\multicolumn{1}{c}{Age of oldest child} &
  \multicolumn{1}{c}{\begin{tabular}[c]{@{}c@{}}Not concerned \\ about breaking device\end{tabular}} &
  \multicolumn{1}{c}{Percentage} &
  \multicolumn{1}{c}{\begin{tabular}[c]{@{}c@{}}No help given \\ with device\end{tabular}} &
  \multicolumn{1}{c}{Percentage} &
  \multicolumn{1}{c}{\begin{tabular}[c]{@{}c@{}}Is competent with\\ the device\end{tabular}} &
  \multicolumn{1}{c}{Percentage} \\ \midrule
Between 0-5 & 19  & 36.5\% & 22  & 55.8\% & 28  & 53.8\% \\
5-10        & 157 & 56.7\% & 183 & 66.1\% & 217 & 78.3\% \\
11-15       & 251 & 66.8\% & 313 & 83.2\% & 297 & 79.0\% \\
16-17       & 136 & 75.1\% & 161 & 89.0\% & 140 & 77.3\% \\ \bottomrule
\end{tabular}%
}

\label{tab:age_oldest_child}
\end{table}

\subsubsection{Cyber Security in the Home}

\paragraph{\textbf{Where do users get support?}}
There was a wide spread of answers to the multiple response question ``Who or where do you turn to for support with digital technology use''. 406 respondents (73.29\% of the total number of respondents) said that they would turn to the Internet for support, with 217 respondents (39.17\%) saying that they would also turn to other adults within the household. 64 respondents (11.55\%) suggested that they needed no support, with similar numbers turning to friends (73, 13.18\%), family outside the household (65, 11.73\%) or children within the home (72, 13.00\%). Very few respondents said that they made use of paid support (6, 1.08\%).

\paragraph{\textbf{Cyber crime and its effects}}
The vast majority of participants had not, or did not believe that they had been, the victim of a cyber crime at any point (457, 82.50\%). Of those that responded that they had, 26 reported having suffered a financial loss as a result. 65 said they had not, with 6 saying that they were not sure. 503 respondents (90.79\%) said that they were unaware of manufacturers of devices that they owned ever having reported a security breach.

\paragraph{\textbf{How is cyber security discussed at school and home?}}
The survey also looked at the parents' understanding of what children learned at school about cyber security and Internet safety. The most common themes were ``cyber bullying'' (411, 74.19\%), ``stranger danger'' (396 respondents, 71.48\%) and ``harmful online content'' (390, 70.40\%). More traditional cyber security themes were much less frequently recorded, e.g., ``use of strong unique passwords'' by 194 (35.02\%) and ``malware'' by 107 respondents (19.31\%). 87 respondents (15.70\%) said that they were unaware of what their children had learned at school. There was a statistical significance between the age of the participant and the knowledge of those types of themes associated with home IoT device use\footnote{With the participant responses being gathered into topics that were home IoT device related (strong password use, data management practices, malware, cyber crime, personally identifiable data), those that are not home IoT device related (stranger danger, harmful online content, cyber bullying) and ``I don't know''.} ($\chi^2(6, N = 552) = 35.99$, $p < .001$) (see Table~\ref{tab:age_participant}), and also the gender of participant and the knowledge of these themes ($\chi^2(2, N = 552) = 12.27$, $p < .002$). The data shows that the older a parent is, the more likely they would be to report such themes; women (44.6\%) were more likely than men (31.7\%) to do so, too.

Finally, the survey reviewed discussions held at home by families about cyber security, and the Internet. 481 respondents (86.82\%) reported having discussed the importance of not talking to strangers online. Other topics were far less uniformly discussed. 245 (44.22\%) had discussed ``what data might identify or reveal about you'', and 212 (38.27\%) had discussed the use of strong and unique passwords. 27.98\% of respondents (155) had had discussions about how to minimize the risks of becoming a victim of malware. There was a statistical significance between the age of participant and the likelihood to discuss topics such as passwords and malware, instead of topics more closely linked to safeguarding, such as stranger danger\footnote{As with the school test described in the previous paragraph, the participant responses being gathered into topics that were home IoT device related, those that are not home IoT device related and ``I don't know''.} ($\chi^2(6, N = 552) = 17.73, p < .007$). The older the age range of a participant is, the more likely they would be to report discussing those topics more closely aligned to cyber security aligned to home IoT device use --- see Table~\ref{tab:age_participant}.

\subsection{Interviews}
\label{sec:interview_findings}

\subsubsection{Summary of findings}

This paragraph gives a high-level overview of the interview findings presented below. Both parents and children talked about threats and risks in the language of online safety --- stranger danger, cyber bullying and so on --- and scams, without association to the home IoT devices that they were discussing.  The ability for children to spend money through home IoT devices was recognized by both parents and children, but often not acted upon until the children had spent money without permission.  Other types of financial fraud were not taken particularly seriously by parents (and was not considered at all by children) as the mechanisms for reimbursement through UK financial institutions were considered to be robust. Data loss from IoT devices was not raised as a concern.

Much as with their description of risks and threats, families described discussing online safety issues, when asked about how cyber security was discussed at home. Evidence of robust cyber security knowledge was relatively limited among both adults and children, with application of that knowledge poorer still. Much stronger concerns arose, from both parents and children, about other aspects of home IoT device use --- not security or privacy, but the costs of the devices, the interoperability capabilities of the devices, and the necessity of careful use to extend out their useful life.

\subsubsection{What do families consider the threats and risks of home IoT device use to be?}

\paragraph{\textbf{Stranger danger, scams: the same threats as posed by using the Internet directly}}

All interviewed families did not differentiate between the threats and risks that home IoT devices pose, compared to other digital technologies that allow you to browse the Internet, and connect to social media in particular. Participants, too, presented no understanding of specific threats based upon their personal situation, but rather focused on generic threats with no clearly defined adversary.

The most talked about threat was that of ``stranger danger'' and that of inappropriate online contact with strangers. This is less applicable to home IoT devices, which participants recognized (``\textit{Well, you can't chat to anyone on the [Amazon Fire] stick.}'' (C5, aged 10)). The second most commonly considered threat was that of financial loss, through falling victim to scams, e.g., ``\textit{... giving your password, your bank password, to someone and they then transfer all your money out. That kind of thing, I suppose. I can't think what else they could do that would bother me that much, or as much as that}'' (AF2). This was a risk that parents felt strongly, and children were much less cognisant of, with suggestion that this was because they were not individually capable of managing their own finances. Interestingly, children who had seen attempted scams were likely to recognize them as such precisely because they targeted accounts they were not old enough to have (``\textit{...I've got like, 10 [text messages] this week [telling me that] my PayPal account has been hacked...but I don't even have an account.}'' (C13, aged 12)).

Some parents felt that they could rely upon being reimbursed, should financial fraud occur, which minimized the harm (``\textit{ ... my PlayStation 4 account was hacked. And they spent a lot of money. I got it all back. It was fine.}'' (AM9)); this was from an expectation that any form of fraud would be reversed, or at least properly investigated, by traditional financial services firms, such as banks and credit card companies (``\textit{ ... [if] someone's in your bank account, you'd call your bank, and they'd resolve it}'' (AM7)). A couple of adult participants had experience of fraud through less traditional financial services firms, such as PayPal:\footnote{PayPal explicitly explains that, because of its choice of location and registration for UK business, ``The nature and extent of consumer protections may differ from those for firms based in the UK.'' \citep{paypal_home_2021}} ``\textit{I just noticed all this abnormal spending activity. PayPal weren't particularly interested, I have to say.}'' (AF8). The participant here was not reimbursed.

\paragraph{\textbf{Unauthorized spending on devices is not as concerning as fraud}}

Unauthorized spending on devices did not appear to be considered ``financial fraud'' for our adult participants, despite being a much more possible threat of home IoT devices. Perhaps as a result of being interviewed during periods of COVID-19 lockdowns, limiting the number of people that could enter the home, any unauthorized spending was considered almost exclusively in terms of children in the home, and as such was considered to be a restriction that parents should manage, even if they were unaware of the ability to do so: ``\textit{Oh yes, ... you ended up spending £5 on some terrible game, because we hadn't yet realized we'd not blocked it ...}'' (AF6). The prevailing feeling, even among the children who discussed it, was that children might be expected to spend money accidentally when using devices, without realising what they were doing (``\textit{I've got a password for my profile...so I don't spend money by accident.}'' C22, aged 8). Those with family accounts on shared devices described instances of receiving an email about online purchases made by the child's account, and having to have a discussion with the child after the event. Only in one instance did a parent give an example of making the child repay the cost of the downloaded goods --- a significant sum was spent on video games, by chance in the run-up to the child's birthday, making paying the parent back with money received as gifts a straightforward event. Had it been any other time of the year, the child would not have had the money to repay the parent.

\paragraph{\textbf{Data loss is not an active concern}} Participants did not exhibit particular concerns about the loss or theft or data of themselves or on behalf of their family from any devices (home IoT or otherwise), or even the ongoing use of personal data by devices and their manufacturers. Some adult participants were sanguine about data loss, e.g., by saying ``\textit{we all just live now assuming that our data has been stolen and [we are] waiting for something bad to happen}'' (AM23). Children, too, exhibited signs that they were aware of the risks of making their personal data available to others, whether through accident or design. In answer to the question ``what is the biggest risk of using home IoT devices?'', some children responded as follows: ``\textit{Your information being shared and people knowing what you do on the Internet...because you could get hacked or something}'' (C10, aged 13); ``\textit{I think, just accidentally giving my information}'' (C12, aged 13), ``\textit{Accidentally sharing information that is going to come back to you and put you at risk}'' (C23, aged 15). Children interviewed under the age of 13 did not talk about data loss as a risk.

Two families with more hands-off devices, such as robot vacuum cleaners, smoke alarms or thermostats, seemed genuinely puzzled by the notion that they would discuss these devices with their children, or that there would be security implications. In both cases, the children were typically reported as being interested and disruptive with the devices for a period, then ignoring them. Upon pressing, one parent explained their thought process about the smart thermostat: ``\textit{I would teach them to use it in the same way I would teach them to use anything else ... as [child] gets older, I would like him to load the dishwasher ...}'' (AF9). When the time came, they would be showing their children only how to use the manual overrides for the smart devices in the home, not expecting them to use apps, or have any more significant control over the management of the data. When asked about data collection from smart home devices, this family felt comfortable with their expected data use by the manufacturers: ``\textit{I know they collect all the information. I know they try and use it to make money.}'' (AM9). Another participant, heavily bought into the Amazon ecosystem of home IoT devices, expressed frustration that the most effective way to limit data collection, or to opt out of data use for improvement, was to activate a setting that might see them not receive future features: ``\textit{I changed some settings in the app that stopped them because they can listen to your recordings ... And you can turn off the ability for them to listen to it. It's like a weird setting because it comes up with an option that says why you might not get any of the new features if you turn this off, and I'm like, `I like new features!'}'' (AM16).

\paragraph{\textbf{Linking ``hackers'' to Internet, not home IoT device use}} Some children considered ``hacking'' to be the biggest risk, but could not necessarily explain what hacking meant, who might hack or what ``being hacked'' might look like in anything more than vague terms, e.g., ``\textit{they can change your password, so you cannot log into anything}'' (C24, aged 8) or ``\textit{to me hacking means all data is stolen then it is sold}'' (C18, aged 10). Children also framed this data capture in terms of smartphone or social media use (``\textit{... phones have, like, tracking information, so they know where you are}'' (C13, aged 12)), rather than home IoT device use. Several parents were unsure whether younger children recognized that home IoT devices connected to the Internet in similar ways to computers or smartphones (``\textit{They don't understand that Alexa is connected to the Internet.}'' (AF20)). When asked, most children were confused by the question of how a home IoT device might work: ``\textit{[Alexa] is a program. So people put in questions and answers. So yeah, it's like Google.}'' (C10, aged 12); ``\textit{It's a good question. I don't know. It's magic!}'' (C23, aged 15). Children seem to have a level of knowledge about data collected when using the Internet directly, but are at risk of not making the link between personal data collection, potential misuse and home IoT devices.
 
\paragraph{\textbf{Threats to the home}} No adult participants raised potential threats to the home from home IoT devices, whether in terms of facilitating burglary or other physical damage. Conversely, two sets of parents explained how their connected home surveillance systems had enabled assistance in arresting a burglar in their neighbourhood (``\textit{We were able to send pictures to the police.}'' (AF2)), and in feeling more secure after past events around the house (``\textit{I think just having it there prevents things.}'' (AF19)). Children seemed to be more aware of physical damage that might arise from devices, and considered these risks more urgently than risks to personal data in two cases: ``\textit{There are a lot of things plugged in down there --- they could set fire!}'' (C19, aged 12); ``\textit{I think the most dangerous thing is getting water inside it...that's the only thing that's dangerous about it.}'' (C5, aged 10). These concerns were not shared by their parents.

\subsubsection{Cyber security knowledge gathering, sharing and adherence within the family}

\paragraph{\textbf{Discussion as a family...}} The most common way that interviewed families reported themselves managing any type of cyber security was through discussing it as a family. Should a child want to use a new app or device, many families expressed a process of discussing guidelines --- length of use at a time, who and how they could communicate with people, what they could download, for example. Despite bringing the topic up when asked about cyber security, families recalled talking more spontaneously about online safety issues that they saw arising from the Internet (posting inappropriate content, not talking to strangers) than more strictly cyber security issues. Whether explicitly understood by the parents or not, the importance of a trusting relationship for this type of discussion was clear: ``\textit{... we talk to them a lot, and tell them to tell us in any situation}'' (AF7). Children often referred to asking or learning from parents when they were unsure (``\textit{I just call Mum and Dad}'' (C7, aged 7), ``\textit{I [watch] my parents do it}'' (C11, aged 14)).

\paragraph{\textbf{...but not about home IoT device security}} There were no reported instances of interviewed families discussing home IoT device security together. Some parents suggested that they might use items in the news as a means of having conversations about specific security topics with their children, although, when asked about what news stories they had seen recently that made them consider their own home IoT setup, no one interviewed pointed to a specific story about a home IoT device being compromised. Neither parents nor children could explain, when asked, in much detail, as to how the devices they used worked: ``\textit{I don't actually know --- no one has ever asked me that question before.}'' (C17, aged 8). Yet parents seemed to assume that at some point there would need to be a discussion, without taking steps to understand the technology further. As AF5 explained: ``\textit{What I'm guilty of is I don't really think about it in any depth. You know, and probably because my children are still fairly young that you don't sort of think any further than what you need to.}'' 

Some parents with a larger age gap between children reflected how they seemed to have missed having conversations with their older, teenage children, because advances in digital technologies had happened relatively recently. Conversely, their younger children seemed to have grown up with, and just ``understood'' core functionalities of devices in the home instinctively, expecting devices to have touch screens and to stream programs on demand, for example, without understanding the technologies behind the device: ``\textit{He knows what he's doing when he's looking at it, but ask him what he's done. He wouldn't be able to tell you.}'' (AM24).

\paragraph{\textbf{Children's awareness of cyber security}} All children recalled that they had had classes at school about certain aspects of computer use. Most talked of having annual reminders of ``stranger danger'' and ``cyber bullying'', and using programs such as \textit{Scratch}\footnote{\url{https://scratch.mit.edu/}} to learn programmatic thinking and the fundamentals of coding. Only one child actively discussed being taught about malware (C23, aged 14). No children reported being taught about IoT products at all, with one child commenting that ``\textit{It would be nice to know just how different technologies worked, not just spending all the time on computers}'' (C2, aged 12). Many children found school's approach ``\textit{all common sense stuff}'' (C7, aged 11), and ``\textit{just talk[ing] about the same thing every time}'' (C5, aged 10). 

Most children interviewed indicated that they knew something about strong passwords, but typically had one core password that they used (``\textit{I have my passwords all the same, so I remember it! But if you hack one account, you hack all of them!}'' (C13, aged 12)), varying them should they need to share details with others (``\textit{... if my friend and I may want to make an account where we all have access, I will use a password that is not my original password.}'' (C5, aged 10)). Beyond this, children exhibited little cyber security knowledge that could be considered specific to home IoT devices: one child had read an article about how to delete their search history from their smart home assistant: ``\textit{I think I probably just saw something about it like on social media: `This is how you stop your Alexa from knowing all your secrets'.}'' (C23, aged 14).

Seven families had instances where a child expressed knowing more than a parent about the Internet generally, or specific security features. In these instances, children pointed out gaps in their parents' knowledge, but gave no indication that they would be expected to teach or help their parents further. For example, C13 (aged 12) showed frustration that his mother ``\textit{didn't understand the safe tick}'' that the anti-malware software they used provided to show the safety of websites. AF13 followed this by saying ``\textit{you'll have to show me!}'', suggesting that the willingness to learn may be present, but not necessarily obvious to the child. Despite the suggestion that children may be on a par with their parents, there was no instance where a family expected the child to have a part in the security or management of communal devices, even if the child had an interest or exhibited accurate knowledge about security.

\paragraph{\textbf{Parents' awareness of cyber security}} Ten of the adults considered themselves to be quite confident in the security of their homes. Either they thought themselves ``\textit{fairly savvy on things}'' (AF1) or felt confident that devices come ready prepared to avoid any kind of security breach ``\textit{Well, I suppose a lot of the devices such as the Echos ... those kinds of things that kind of come pre-set up to be defensive}'' (AM2). AM16 was aware of the potential for insecure home IoT devices to be used as a means of getting access to an entire network, but put faith in well-known devices: ``\textit{I read something about someone who put a Wi-Fi kettle on their network, and then someone had hacked into it ... and gained access to the whole [network] ... It can be a bit dodgy, but I trust the Amazon system, the Alexas ...}'' Participants that were confident of their abilities to secure computers did not provide evidence of having more security measures set up for smart home devices (for example, discussing guest networks, non-standard router settings or additional home networking) than users who were not confident when asked. Almost all adult users, and some of the older children, suggested that, should they need to find information about any aspect of security, they would search online, and assess the responses based on their judgement or prior experiences: ``\textit{Slight clich\'e, but Google, Google, and then I look at what looks like a legitimate organization. I don't have like, a specific way in mind, but I think I'm able to tell based on you know, how the website looks.}'' (AF9). 

When adults were asked how they would like to receive information about cyber security, answers ranged from getting it from trusted technology TV shows, popular UK consumer champions, private companies, or even school. There were some parents who thought that this might be the remit of the government, but were unclear where it might sit (Interviewer:``\textit{Which areas of the government?}'' AF25: ``\textit{I would say, probably [the Department for] Education ...}''), or if there was even sufficient budget for the government to do it usefully (``\textit{...maybe the government but I don't know how the government actually have the resources.}'' (AF8)). AM3 reflected that it would be helpful if there was an ``\textit{NHS} (National Health Service) \textit{for cyber security}''. The UK'S NCSC, who run the national security awareness campaign CyberAware,\footnote{\url{https://www.ncsc.gov.uk/cyberaware/}} was not mentioned by any participants.

\paragraph{\textbf{Parents' cyber security application}} Several adult participants expressed their inability to understand what is needed to secure the home and act accordingly, relying on other adults in the home (``\textit{My husband brings stuff in the house and I have no clue. I can't tell you exactly what we've got ... I've got no clue how it works or anything about it.}'' (AF2)), or not thinking about it (``\textit{I'm one of those people who probably doesn't think about security. I just go along until the disaster strikes and then yeah, I mean, obviously, I've got you know, passwords for our various things, but I mean, I don't sort of you know, think about it in that much depth.}'' (AF5). Three adult participants equated being secure with having anti-malware software (``\textit{Once I download it I am happy for the year.}'' (AF13)) --- even when not being sure that it would carry over to the home IoT devices they had: ``\textit{... [using anti-malware software] was just kind of what got drummed into us when we were younger and when we first had computers, but with all the other kind of connected items, you just assume it's ... in there, I guess.}'' (AF6). Other participants understood that anti-malware was not the sole means of keeping their home secure, and was not on their home IoT devices: ``\textit{I have Norton on everything apart from the TV.}'' (AF 17). This gap was noted as if the home IoT device did not require any additional protection. Some participants considered free anti-malware to be superior to paid-for versions (``\textit{... I found a lot of them if you'd shop around the free ones are just as safe as the subscription ones. I mean, I swear on AVG Free for most of it. I still swear with that over £60 Norton every time.}'' (AM24)). Given the practice of selling the data of users of free antivirus software, this may, in fact, open the user up to different security issues \cite{soni_avast_2020}.

\paragraph{\textbf{Where cyber security knowledge was lacking}}

Only one interviewed adult participant used a password manager for themselves, with another participant writing passwords down in a notebook. Two other participants expressed awareness of password management software, but dismissed them as too complex and tricky, or incompatible with not having access to technology all the time (one participant, unable to use their smartphone at work, was unable to access their bank account because the details were stored in the password manager). Other adult participants did not consider password management outside of their own memory a necessary practice. Some were confident that they could store a range of passwords adequately without further assistance (``\textit{I'm fairly on the ball myself with it, to be honest, so I don't ever see the need [for a password manager]}'' (AM24)), many expressed some small discomfort, or guilt, at knowing that they did not follow guidance that they knew to be important (``\textit{I have the same passwords pretty much. Just like a couple of letters changed. I'm terrible, yeah I know ...}'' (AF21)). Those adults participants asked had not changed the password on their router (``\textit{It's too bloody long!}'' (AM7); ``\textit{... I just put that in the too boring/too hard basket.}'' (AF1)), and retained the router given to them by their Internet Service Provider. Children did not reference using password management software themselves, or having any discussion with their families about how to manage passwords of home IoT devices, and the potential for misuse when a single account is tied to a communal device.

Few interview participants, and no children, reported using multi-factor authentication (MFA). In the majority of cases where individuals did report using MFA, it was only done after being the victim of a cyber crime, resulting in fraudulent activity (``\textit{[the experience] made me go and turn on 2FA}'' (AM9)). MFA is, of course, less obviously used in home IoT devices directly, however equivalent steps, such as requiring a PIN for purchases through an account linked to a home IoT device, were typically not activated by participants. As with password management, there was no report of discussions within the family about the use and benefits of MFA.

Most adult participants were not completely sure as to whether their devices received automatic software updates, or whether they needed to be involved in the process. There were no participants who were aware of how long they expected their devices to be supported for, with some participants misunderstanding the question (``\textit{[supported life] is probably similar to a normal energy bulb}'' (AF16)). Software updating was not discussed between families as a necessary process. Three children mentioned the importance of automatic software updates --- but only one in relation to a device other than their smartphone (an Amazon Fire Stick). Six adult participants talked of their assumption that updates were automatically done, and took little time to consider it: ``\textit{when they get notifications they'll do it, but it's not a conscious thing.}'' (AF14, about the children in the house). Both adults and children expressed a desire to manage their device updates, either because they did not want to have their use interfered with in the moment (``\textit{I feel like it's an informed choice when I do things}'' (AF2)) or because ``\textit{I feel like it slows the phone down}'' (C25, aged 16).

\paragraph{\textbf{Other home IoT concerns: costs, interoperability}}
A much larger concern than security, particularly for those with a larger number of home IoT devices, was how to expand their smart home within their budget, and keep devices functional for as long as possible. Some participants were proud that they had been careful with old smart home assistants that were still functioning: ``\textit{ ... none of [my Echos] have ever broken. Some of them I've had since right at the beginning.}'' (AF19). One participant with a significant smart home set up recalled that, with individual devices being so expensive, building up a smart home system was a slow process involving second-hand devices and constant interoperability issues. Older devices that this family had did not have full functionality, compared to newer devices, suggesting being out of supported life. Another family, unaware of the security implications, talked at length about how these interoperability issues made life a little more frustrating and a little less smart than they would have liked (``\textit{Having to have a bridge like that is really annoying...why does Phillips [manufacturer of smart bulbs] need a bridge [to work with other home IoT devices] when none of the others do?}'' (AF19)). 

There were agreed-upon strategies to ensure that children took care of their devices --- some children recognized that a device broken by them would not be immediately replaced (``\textit{I don't want my [device] to break, because I won't be getting another one!}'' (C6, aged 6)), where other families employed strategies such as this: ``\textit{We make them spend their own money on stuff like that. Hopefully promote a sense of like, you know, `I bought this, it’s mine and it's my responsibility.' ''} (AF7). Younger children in particular exhibited signs of friendship and care towards smart home assistants (``\textit{[C4. aged 4] considers the Google Home part of the family...they have conversations in a very friendly way.}'' (AM4)), or seeing them as integral (``\textit{just part of our house}'' (AM16)).

\section{Discussion}
\label{sec:discussion}

\subsection{A misalignment of perceived and actual threats and risks}

\subsubsection{Perceived threats and risks}

The interviews and the survey showed that families were happy to use and integrate home IoT devices in their homes, and did not take particular steps to inform themselves as to how to make these devices more secure. The underlying concerns of parents in particular, around stranger danger and financial fraud, suggest that they did not recognize the difference in threats and risks posed by home IoT devices, compared to using the Internet in a home computer. Children echoed this, speaking almost exclusively of threats posed to them on social media, and risks typically stemming from ``hackers'', external to the family. 

The adults and children interviewed had some level of understanding of how having home IoT devices increased the collection of personal data, and that misuse of such data by someone with malicious intent could be bad. {The survey showed a complex relationship between the age of the participant and their actions in relation to their children --- regardless of the child's age: although, the older the parent, the more likely they were to report talking to their children about cyber security measures that could be applied to home IoT devices, they were much less likely to offer their children support with such devices. This shift perhaps reflects the adoption of digital technologies over time --- younger parents, more likely to have experience of digital technologies as ubiquitous and for interacting with others, seem to focus more on the risks arising from this.} The interviews and survey indicated several reasons for the lack of knowledge and subsequent action by participants. As indicated by the survey results, people did not notice news stories about their devices, and may miss stories of data leaks and other breaches, thus not taking actions when needed. The interviews highlighted how adults and older children were resigned to data being taken and occasionally leaked, and the survey results suggest that people are often unaware of, or immediately unaffected by any breach. Regardless of their feelings about this, if individuals want to use such devices, they have to accept the terms of use, which often requires full data collection to be turned on for complete device functionality. Furthermore, cutting corners with, or not using, cyber security measures rarely causes directly applicable, irreversible harms, which reduces the incentive to try to change.

\subsubsection{Actual threats and risks}

There is the overwhelming feeling that threats posed by home IoT devices are too impersonal and too remote for many people to feel compelled to take interest. It has long been considered that individuals do not find themselves interesting enough --- or not wealthy or valuable enough --- to be hacked \citep{howe_psychology_2012}; in this light, it is understandable that the families interviewed spent more time discussing the much more manageable and realistic risks of talking to strangers and over-sharing personal data online.

It also does not help that stories, such as the home network being hacked through an insecure smart kettle, as mentioned by one of the interviewees, have a faintly ridiculous edge that undercuts the severity of having a home network compromised. Conversely, other potential threats that can arise from a malicious party accessing the types of data that can emanate from a smart home can be extremely distressing to consider, given the potential of harm to property and person, and how little the user might be able to do to protect the household. The 2019 incident of an individual hacking into a Ring camera in a child's bedroom and pretending to be Santa, talking directly to the children \citep{noor_ring_2019}, is one such upsetting example. Not only is this reported to be an attack widely possible due to software especially designed and distributed for the purpose \cite{cox_ring_2019}, but Amazon, the owner of the Ring device family, has refused to acknowledge their responsibility for the harm and distressed caused, blaming users' insecure practices for facilitating the hacking \citep{paul_ring_2020}. The psychological harms of an event like this need to be appropriately considered when considering threats, particularly when taking decisions for, and discussing alongside, children or other dependents. \citet{agrafiotis_taxonomy_2018}, considered the range of potential psychological harms that may occur to victims of a cyber breach --- ranging from embarrassment and discomfort, to shame, depression and guilt, over an extended time period. Within a family setting, these types of feelings can be difficult to discuss, or may affect different family members in different ways. The importance of effective cyber security measures --- strong and unique passwords, and regularly updated software in particular --- is even more important when manufacturers may cite a user's lack of management of the security features to avoid liability.

Children interviewed did not report witnessing active risk management from their parents. The survey results showed the extent to which users did not know the details of their devices' security features. The interviews showed that families did not routinely talk about cyber security strategies that home IoT devices may need to use, such as the deploying of password management strategies, using MFA or the importance of regular software updates. Additionally, women were less likely to manage the devices in the home, or understand how to use them, but take more ownership of understanding what their children learn at school, meaning that the parent that may be less capable or confident is the one overseeing the children's education in relation to applicable cyber security. Instead, children are taught the importance of managing the cost of device use from their parents. Parents may look to second-hand devices to reduce the cost of expanding their smart homes, agree with their children that they must take care of those devices that they use, and use free security software. It is a common practice that software is free at point of use, but is monetized through strategies such as selling data collected about the users. This can be extremely invasive when software is designed to control an entire system, as is the case with anti-malware applications \citep{soni_avast_2020, ekambaranthan_money_2021}. Additionally, normalizing the downloading of free software may increase the chance of downloading a malicious app \citep{vaas_android_2020}; although directly a concern on smartphones, similar threats from careless downloading of skills (the smart home assistant equivalent of apps on mobile devices) have been shown to exist \citep{lentzsch_hey_2021}.

Parents promoted ideas of relying upon recognized consumer protection regulation, particularly in the case of fraud, but also in ignoring children who believed electrical safety could be an issue. Adults in the UK are accustomed to a rigorous consumer standards framework, with the ability to approach financial services firms for reimbursement where there is fraud without ``gross negilgence'' \cite{ombudsman_fraud_2021}, and levels of assurance around electrical safety.\footnote{Such standards include the Kitemark (\url{https://www.bsigroup.com/en-GB/kitemark/}) and the CE mark (\url{https://www.gov.uk/guidance/ce-marking}).} Families also modeled behavior about extending device life through careful use. In a family context, this poses the problem that devices are not necessarily as secure as people might expect, leaving them open to attack in ways that may not be covered by consumer protection regulation, yet remain damaging to those affected.

\subsection{Neither adults nor children have opportunities to learn appropriate cyber security information}

Both the survey and the interviews show the lack of appropriate education opportunities for both adults and children around cyber security. Children are more easily approached in school settings, but the status quo in the curricula applicable in the United Kingdom is not teaching appropriate security for home IoT devices or capturing the interest of children. Those children interviewed found cyber education at school repetitive, learning little that they did not already know about some topics (notably the safeguarding aspects of Internet use), and little to nothing about security topics. This confirms the findings of \citet{Pencheva_bringing_2020}, focusing on cyber security at school, who found that students generally are very tech-savvy, but treat online safety and cyber security as the same subject. The survey, furthermore, showed how important it is to ensure all children in a family receive the same education in a school setting, as it appears to be the case that the presence of older children in a home setting reduces the amount of attention given by parents to ensuring careful home IoT device use and secure cyber security practices of all the children.

Education for adults in this space, outside of formalized education, is extremely difficult to land effectively \citep{bada_cyber_2019}, and the efficacy of such education depends upon personal and social norms, in many respects. Our interviews did not get as far as asking what types of message would be personally effective for participants, given the basic lack of knowledge as to how the devices worked on the part of the participants. However, participants did give a suggestion of the types of organisations and groups that they would like to hear from. These ranged from the government and related bodies, to trusted people with influence (such as particular TV shows and consumer champions) or manufacturers. The reliance on turning to the Internet for advice shows a lack of knowledge that could possibly be bolstered by these channels. Of course, this depends on how well trusted organizations are, which is a factor that will change from country to country, as will the societal approach to parenting, and managing personal risks. As an example, \citet{kritzinger_study_2017} found that, in comparing cyber security message approaches between the UK and South Africa, the UK's messaging focused upon the responsibility of the individual to protect themselves, rather than being a communal effort. This ties in with the argument in \citet{renaud_is_2018} that cyber security in the UK is becoming ``responsibilized'', with individuals expected to manage their own risks. This is not a particularly unusual expectation, also seen in the USA \citep{haney_it_2021}, but, as the interviews showed, if individuals have inaccurate understandings of where and how they are protected, and what the actual threats and risks are, it is extremely unlikely that they will manage these risks effectively.

\subsection{Recommendations to help families become more secure in their home IoT device use}

\subsubsection{Help families understand where to do better}

Without even a general knowledge, it is hard to take the next step of efficient threat management based upon personal situations. Extending the suggestions from \citet{parkin_security_2019} and \citet{emami-naeini_exploring_2019}, manufacturers must do more to make the key threats of home IoT device use clearer at point of setup, and periodically throughout device use. Documentation about major threats relating to personal data and device misuse, and the recommended baseline steps to mitigate these issues based upon the specific device may help to increase awareness in adult users, which can, in turn, trickle down to children. Manufacturers may also focus on providing this information clearly online, so that when users are faced with trying to improve their knowledge, official sources are plentiful.

Home IoT devices --- particularly those that must have a reasonable expectation of use by multiple people, multiple generations, within a single location --- should be provided with better guidance as to how to set them up for the situation it is located in. Every person, and every household has its own particular set of threats and risks that might make certain security options more or less appropriate for them; however, there will be certain baseline security features that should be promoted to users at the point of setup and periodically thereafter. \citet{hiniker_not_2016} and \citet{ko_familync_2015} have shown that families are more likely to come to agreements over rules for Internet use if they agree upon these rules as a unit. Our findings agree with this: families instinctively use discussion as a means of managing potentially tricky areas for their children. However, previous research has shown children and adults may use different words to discuss cyber security \citep{jones_what_2019}. In addition, children may not perceive devices in the same way that adults do, not only, as our interviews showed, failing to understand they are connected to the Internet, but also in understanding how they work \citep{yip_laughing_2019,xu_what_2020}. As such, the expectation that using discussion either to agree about home rules about home IoT devices or as a means of providing children with additional knowledge will fail, without further guidance for both parents and children. 

\subsubsection{Appropriate levels of base security measures for home IoT devices}

The evidence from the interviews in particular suggest that home IoT devices that are expected to be used by groups of people --- especially children --- should come with more stringent security measures set by default. For example, although some steps seem to have been taken by manufacturers to allow notifications when things are purchased, further work could make spending money and downloading items harder (requiring purchases to be confirmed with a PIN or other agreement from the primary account holder). Alternatively, a clear signpost to the potential risks of not protecting purchases at the point of setup, should be provided. The interviews show that, for example, adults in the UK are used to being protected by strong consumer protection measures that should cover most cases of fraud --- but crucially, not ``gross negligence'' when, despite being the victims of fraud, they were deemed not to have taken the proper precautions \citep{brignall_bank_2020}. Although adult interview participants were extremely concerned about financial fraud, few had enabled settings to limit purchases on applicable home IoT devices, and no family considered the ease with which significant sums of money could be spent fraudulently, potentially without recourse.

It must be recognized that there are proposed measures to mitigate some of the security aspects explored here, such as the UK's Product Security and Telecommunications Infrastructure Bill, mandating some aspects of security by design in consumer smart products \citep{dcms_bill_2022} and ETSI's EN 303 645 baseline standard for consumer IoT \citep{ETSI_standards_2020}.  Much like the devices they seek to legislate and standardize, these measures are nascent, and may well need to be extended and revised as the market matures. Policy makers should continue to analyze the ongoing harms arising from the increased ubiquity of such devices to understand whether more regulation or legislation may be necessary to protect consumers and specific groups of vulnerable people in the future. An example of regulatory focus on a particularly vulnerable group would be the UK's Age Appropriate Design Code \citep{ico_age_2020}, which requires organizations to consider the safety of children explicitly, when providing those services within the UK.  Acting unilaterally, as the UK government is in the case of \citep{dcms_consultation_2020} and \citep{ico_age_2020}, works on the assumption that the measures can be applied to --- and enforced against --- those devices that are for sale in the UK in a meaningful manner, despite home IoT devices being made, and sold, globally. Some small shifts have been seen in large online platforms' age confirmation processes as a result of the Age Appropriate Design Code \citep{hern_techscape_2021}, but what the effect on the device market will be remains to be seen. 

\subsubsection{Up to date education --- for children and adults} 

Parents and children interviewed and surveyed in this research portrayed the learning provided by school as focusing only on the Internet as a place for socialising and commerce. Curricula must be updated to capture not only the pervasive nature of the IoT, including exploring how the data collection from such devices is even more pervasive than that collected using the Internet, and also why the threats are different to the Internet accessed via a computer. Helping children be able to better comprehend what and who may actually constitute a threat, and how to apply knowledge about cyber security skills to manage such threats would be an excellent start.

However, enhancing school curricula is not enough to improve the cyber security knowledge within the home in the short term --- not only do children not typically have the power in the familial relationship to engender change of their parents' activities, but the accounts for the home IoT devices are unlikely to be theirs to manage. As \citet{blum-ross_parenting_2020} considered, the information exchange between school and parents is poor: there can be no expectation that parents would become significantly more knowledgeable simply because their children are learning more. Parents need to find means of educating themselves, and knowing how to extend this knowledge to their children, where the knowledge is not gained at school.

In the UK there is some expectation of support from the government for personal risk management. However, the interviewees showed a complete lack of awareness of the existence of the body tasked with supporting them with cyber security, the NCSC. There is an interesting case to be made for learning the public health messaging lessons deployed to spread messages around social distancing, and then the promotion of vaccination, in relation to COVID-19. The use of poster campaigns, short slogans and promotion of positive behavior by trusted members of society, whether in a personal capacity or through platforms such as TV shows or recognized websites may serve to normalize good cyber security practices. This should sit on top of school-based education, and would serve to provide knowledge that is currently unavailable to both adults and children alike.

\subsubsection{Better recognition that families try to look after their devices}

The importance of strong security practices becomes more important as countries, including the UK and the European Union, introduce rights to repair in law for certain home electronic goods, including televisions and refrigerators \citep{harrabin_right_2021}. As these devices are expected to be repairable for up to ten years, so smart instances of these devices need to maintain appropriate software updates and usable security processes for this period. This must happen to ensure that users are not put at risk for being careful with their devices, and that children do not need to learn that the ability to have devices on a budget does not come at the expense of its, and their, security.

\section{Limitations and Future Work}
\label{sec:limitations_future_work}

As often happens in research relating to the home and family life, there was an over-representation of women participating in both the survey and the interviews. Although this may serve to skew some responses, as women may have a tendency to underplay their technology knowledge \citep{rode_roles_2010}, this is to some degree balanced by the input of other family members in the interview process. Additionally, although there is clearly going to be participant bias in terms of the type of person who would consider answering surveys on Prolific, this does not seem to have produced a respondent group who is extremely technologically aware, for example. It must be considered also that the research is focusing on technology that is, at present, largely discretionary in nature; as such the higher proportion of employed participants may not be entirely surprising.

As is the case with most research based around qualitative work, our findings can be said to be applicable to the situation that families found themselves in the UK, at the time of undertaking the research. Parenting methods, trust in government, governmental entities and consumer protections vary throughout the world: future work could be undertaken to understand where the findings of this piece of research may differ in other jurisdictions and regions globally. Taking the EU Kids Online survey as an example, which in its 2020 review looked at online behaviors of children in 19 EU countries \citep{smahel_EU_2020}, it may even be beneficial to scale such research up to look at the security differences in home IoT devices between countries: as devices are sold globally, so their use, and misuse, need to be understood globally too.

Future work should build on these findings as a means of engaging families more with the elements of cyber security they can discuss and manage appropriately as a specific multi-user case. In particular, work should look not only at what families can be expected to do, but what areas they do not consider significantly threatening enough to spend time managing, and which bodies should be managing such risks. It is also necessary for further work to be done to integrate knowledge of home IoT devices and the cyber security risks that they pose, into school curricula, to make cyber security education more than an online extension of child safeguarding principles.

\section{Conclusions}
\label{sec:conclusions}

This research looked to answer two initial research questions: how families in the UK manage the cyber security of their home IoT device use, and how and when they learn and discuss cyber security for home IoT device use. We found that families have limited knowledge around the threats and risks of home IoT device use, falling back on more general knowledge about the risks of using home computers. Furthermore, parents showed much greater concern for modeling and teaching their children efficient and careful device use to extend the device's life, in part because of the expectation of harm reduction and management through consumer protection regulation in the UK. We also found that children do not receive any education about the concept of the IoT, or cyber security as it should be applied to ubiquitous devices, as part of the school curriculum. These findings pose immediate concerns, in that home IoT devices do require different cyber security management than home computer use; in particular, the expectation of long device use may see families using unsupported, therefore insecure, devices, and the lack of friction around downloads and purchases especially may not be covered by consumer protection legislation as the interview participants may assume. This also has the potential to store up problems for the future: with no education from either home or school, children may grow up without a proper appreciation of how to keep home IoT devices safe.

We discuss how the UK has a relatively individualized approach to personal risk management, albeit with an understanding that governmental agencies and legislation will protect consumers from most active threats. Specifics will change, based upon where citizens of any particular jurisdiction place their trust, but increased messaging from trusted local sources must be leveraged to improve adult knowledge, on top of incorporation of learning about the IoT, and effective cyber security measures, during a child's school life. In addition, manufacturers must present more details as to the threats, and the specific mitigation factors for their devices, so that, even in the face of a lack of education from the state, individuals can make decisions that are truly representative of the threats and risks to their own families.

\bibliographystyle{ACM-Reference-Format}
\bibliography{main}

\clearpage
\appendix 
\section{Interview questions}
\label{appendix_interview_questions}
\subsection{Device questions}

\subsubsection{To each primary school aged child (4-11):}
\begin{itemize}
\item How do you use the device?
\item Do your parents have to help you?
\item Do you have your own account? If so, does that limit what you can do with it? Do your parents control what you can do (e.g. with an app)?
\item Do you do different things with the device to the adults in the house or your siblings?
\item Can you tell me how you think the device works?
\item What do you like about the device?
\item What would you change about the device?
\item Do you know how to turn the device on or off, or make changes to how it works? Have you tried to do this?
\end{itemize}
 
 \subsubsection{To each 12-16 year old child:}
 \begin{itemize}
\item How do you use the device?
\item Do you have a separate account to other house members? Do your parents have to monitor what you do with the account?
\item How do you think the device works?
\item What information do you think the device has about you? 
    \begin{itemize}
        \item How do you feel about that?
        \item Who do you think can access that information?
    \end{itemize}
\item Do you think you use it differently to your parents and siblings?
\item Have you ever tried to change the settings of the device? If so, can you tell me more about that?
\item Do you think the device is useful?
\item What would you change about the device?
\item Does the device stop you from doing things, physically or virtually?
\end{itemize}

\subsubsection{To each adult (+16) household member individually:}

\begin{itemize}
\item How do you use the device?
\item Do you have a separate account to other house members?
\item Have you ever tried to change the settings of the device? If so, can you tell me more about that?
\item What information do you think the device has about you? 
    \begin{itemize}
        \item How do you feel about that?
        \item Who do you think can access that information?
    \end{itemize}
\item Do you think the device is useful?
\item Do you use the device for the purpose that you bought it?
\item What would you change about the device?
\item Do you have to use workarounds to use the device in the way you want to?
\item Does the device work as well for every member of the household?
\item Do you intend on buying or having any more devices installed in your home in the next year?
\item If so, which and why?
\end{itemize}

\subsection{Cyber security knowledge}

\subsubsection{Questions to all children:}

\begin{itemize}
\item What have you learned at school about computers and the Internet?
    \begin{itemize}
        \item Can you describe to me what you know about staying safe when using the Internet?
        \item Do you talk about what you’ve learned with your family?
    \end{itemize}
\end{itemize}

\subsubsection{Questions to 12+ children:}

\begin{itemize}
\item How much knowledge do you think you have about using the Internet compared to your parents?
    \begin{itemize}
        \item Have you ever given them help?
    \end{itemize}
\item Could you describe to me what you think the biggest risks of using Internet-connected devices in the house are?
    \begin{itemize}
        \item What changes would you like to make to stop those risks happening? 
    \end{itemize}
\end{itemize}

\subsubsection{Questions to adults:}

\begin{itemize}
\item How well do you think you have secured the devices you have in your home? 
    \begin{itemize}
        \item Can you give details?
        \item Is there one person in the house that is more likely to manage issues with technology?
            \begin{itemize}
                \item Is this because of necessity or choice?
                \item Do the other members of the household wish they could do more?
            \end{itemize}
    \end{itemize}
\item What do you consider to be the biggest risk that you have with the IoT devices in your home?
    \begin{itemize}
        \item Is there anything you feel you could do immediately to minimize that risk?
        \item What else would you like to see done to minimize that risk? By whom?
    \end{itemize}
\item Are you aware of ever having been the victim of a cyber crime or data breach?
\item Who manages the cyber security aspects of the devices that you use in the home?
    \begin{itemize}
        \item Is that an intentional choice?
    \end{itemize}
\item Which sources to do you use to get cyber security information?
    \begin{itemize}
        \item Are you aware of educational schemes run by the government or other non-govermental organisations? If so, have you used them, and did they help?
    \end{itemize}
\item What would be your ideal method of becoming informed about cyber security issues that might affect you?
\item Are you aware of any issues (on the news or otherwise) that have been raised about the security of IoT devices?
    \begin{itemize}
        \item How does that make you feel?
    \end{itemize}
\end{itemize}

\clearpage
\begin{landscape}
\section{Code book}
\label{appendix_code_book}

\label{tab:code_book}

\begin{longtable}{@{}ll@{}}
\hline
\multicolumn{1}{c}{\textbf{Codes/subcodes}}                                              & \multicolumn{1}{c}{\textbf{Description of subcode heading}}                                                                                               \\ 
\hline
\endfirsthead
\hline
\multicolumn{1}{c}{\textbf{Codes/subcodes}}                                              & \multicolumn{1}{c}{\textbf{Description of subcode heading}} \\
\hline
\endhead
\hline \multicolumn{1}{r}{\textit{Continued on next page}}\\
\endfoot
\hline
\endlastfoot

{\textit{\textbf{Use of devices within the home}}}                                   &                                                                                                                                                            \\ 
\textbf{Family relationship with devices}                                                & \begin{tabular}[c]{@{}l@{}}Elements of the family's relationship \\ with home IoT devices\end{tabular}                                                     \\ 
Limitation of  use                                                                       &                                                                                                                                                            \\
Lack of trust                                                                            &                                                                                                                                                            \\
Negative managing experiences                                                            &                                                                                                                                                            \\
Positive/neutral managing experiences                                                    &                                                                                                                                                            \\
Controls that people have                                                                &                                                                                                                                                            \\
Negative device experiences                                                              &                                                                                                                                                            \\
Positive device experiences                                                              &                                                                                                                                                            \\
Joint device use                                                                         &                                                                                                                                                            \\ \hline
\textbf{Adult relationship with devices}                                                 & \begin{tabular}[c]{@{}l@{}}Specific instances of how adults use home \\ IoT devices, as described by both adults and children\end{tabular}                 \\
Adult device use                                                                         &                                                                                                                                                            \\ \hline
\textbf{Child relationship with devices}                                                 & \begin{tabular}[c]{@{}l@{}}Specific instances of how children use home \\ IoT devices, as described by both adults and children\end{tabular}               \\
Overly comfortable with devices                                                          &                                                                                                                                                            \\
Child device use                                                                         &                                                                                                                                                            \\
Child ease of use                                                                        &                                                                                                                                                            \\
Child managing or altering device                                                        &                                                                                                                                                            \\
Child circumvents restrictions                                                           &                                                                                                                                                            \\
Child is not careful with devices                                                        &                                                                                                                                                            \\
Child is careful with devices                                                            &                                                                                                                                                            \\ \hline
\textbf{Confusion over device knowledge}                                                 & \begin{tabular}[c]{@{}l@{}}Examples of where any interview participant \\ expressed a lack of knowledge about their home IoT devices\end{tabular}          \\ 
Don't understand how device works                                                        &                                                                                                                                                            \\
Lack of personal confidence                                                              &                                                                                                                                                            \\
Incompatibility                                                                          &                                                                                                                                                            \\\hline
\textbf{How do devices work}                                                             & \begin{tabular}[c]{@{}l@{}}Topics that came up when any interview participant \\ explained their understanding of how home IoT devices worked\end{tabular} \\ 
Data devices have                                                                        &                                                                                                                                                            \\
Always listening?                                                                        &                                                                                                                                                            \\
Never considered                                                                         &                                                                                                                                                            \\
Not sure how secure it is                                                                &                                                                                                                                                            \\
Comfortable about use in home                                                            &                                                                                                                                                            \\
Not comfortable about use in home                                                        &                                                                                                                                                            \\
How much can they do with the data?                                                      &                                                                                                                                                            \\
Download of personal data                                                                &                                                                                                                                                            \\\hline
\textbf{About the devices}                                                               & \begin{tabular}[c]{@{}l@{}}Explanations of how interview participants \\ viewed the devices and why they bought them\end{tabular}                          \\ 
Just like any other white good                                                           &                                                                                                                                                            \\
Why bought?                                                                              &                                                                                                                                                            \\ \hline
\textbf{Parental relationship with devices}                                              & \begin{tabular}[c]{@{}l@{}}How parents view home IoT devices \\ with respect to their children's use\end{tabular}                                          \\
\begin{tabular}[c]{@{}l@{}}Devices aren't different\\ to other technologies\end{tabular} &                                                                                                                                                            \\
Content concerns                                                                         &                                                                                                                                                            \\
Destruction of device                                                                    &                                                                                                                                                            \\
Kid's filter is too restrictive                                                          &                                                                                                                                                            \\
Concerns for children                                                                    &                                                                                                                                                            \\
Educating yourself to look after children                                                &                                                                                                                                                            \\ \hline
\textbf{Where to get support from}                                                       & \begin{tabular}[c]{@{}l@{}}Where participants explained they would \\ get information about security, should they need it\end{tabular}                     \\
TV programs                                                                            &                                                                                                                                                            \\
Online - specific source                                                                 &                                                                                                                                                            \\
YouTube                                                                                  &                                                                                                                                                            \\
Bank                                                                                     &                                                                                                                                                            \\
Government                                                                               &                                                                                                                                                            \\
No idea                                                                                  &                                                                                                                                                            \\
Police                                                                                   &                                                                                                                                                            \\
Family                                                                                   &                                                                                                                                                            \\
Friends                                                                                  &                                                                                                                                                            \\
Online - generic                                                                         &                                                                                                                                                            \\ \hline
\textbf{Concerns}                                                                        & \begin{tabular}[c]{@{}l@{}}All concerns listed, by any participant, \\ about digital technology use\end{tabular}                                           \\
Privacy                                                                                  &                                                                                                                                                            \\
Security                                                                                 &                                                                                                                                                            \\
Unintended consequences                                                                  &                                                                                                                                                            \\
Parental tracking                                                                        &                                                                                                                                                            \\
Lack of parental control                                                                 &                                                                                                                                                            \\
Restrictions                                                                             &                                                                                                                                                            \\
Children believing everything online                                                     &                                                                                                                                                            \\
Child/restricted accounts                                                                &                                                                                                                                                            \\ \hline
\textbf{Risks}                                                                           & \begin{tabular}[c]{@{}l@{}}The risks that participants saw in \\ relation to using home IoT devices\end{tabular}                                           \\
Data breach                                                                              &                                                                                                                                                            \\
Time waste                                                                               &                                                                                                                                                            \\
Physical damage                                                                          &                                                                                                                                                            \\
Malfunction                                                                              &                                                                                                                                                            \\
Inappropriate content                                                                    &                                                                                                                                                            \\
Viruses                                                                                  &                                                                                                                                                            \\
Lack of accountability for online info                                                   &                                                                                                                                                            \\
Identity Theft (real or accidental)                                                      &                                                                                                                                                            \\
Unsolicited contact from strangers                                                       &                                                                                                                                                            \\
Scams/financial fraud                                                                    &                                                                                                                                                            \\
"Hacking"                                                                                &                                                                                                                                                            \\
Doxxing/identifying yourself                                                             &                                                                                                                                                            \\
Information theft                                                                        &                                                                                                                                                            \\ \hline
\textbf{Parent/child discussion}                                                         & Types of family discussions described by participants                                                                                                      \\
Discussing things learned at school                                                      &                                                                                                                                                            \\
Discussion as a means of solving problem                                                 &                                                                                                                                                            \\
Children learn from parents                                                              &                                                                                                                                                            \\
Safeguarding                                                                             &                                                                                                                                                            \\
Child telling parent what to do                                                          &                                                                                                                                                            \\ \hline 
{\textit{\textbf{Cyber security management in the home}}}                            &                                                                                                                                                            \\
\textbf{Knowledge}                                                                       & \begin{tabular}[c]{@{}l@{}}Instances in which the participant's level of, \\ or means of acquiring, knowledge were explained\end{tabular}                  \\
Tech negatively in the news                                                              &                                                                                                                                                            \\
Tech in the news                                                                         &                                                                                                                                                            \\
How to do it                                                                             &                                                                                                                                                            \\
Not that concerned                                                                       &                                                                                                                                                            \\
Self-confidence                                                                          &                                                                                                                                                            \\
Too difficult to think about                                                             &                                                                                                                                                            \\
Good level of security in the house                                                      &                                                                                                                                                            \\
Unsure of level of security in the house                                                 &                                                                                                                                                            \\
What is taught at school                                                                 &                                                                                                                                                            \\
What school could cover                                                                  &                                                                                                                                                            \\ \hline
\textbf{Strategies}                                                                      & Examples of specific cyber security strategies                                                                                                             \\
Turning off WiFi                                                                         &                                                                                                                                                            \\
Password manager                                                                         &                                                                                                                                                            \\
Awareness of malicious things online                                                     &                                                                                                                                                            \\
Smishing                                                                                 &                                                                                                                                                            \\
Old devices                                                                              &                                                                                                                                                            \\
Biometrics                                                                               &                                                                                                                                                            \\
2FA                                                                                      &                                                                                                                                                            \\
Data deletion                                                                            &                                                                                                                                                            \\
Anti-malware                                                                             &                                                                                                                                                            \\
Password reuse/weak passwords                                                            &                                                                                                                                                            \\
Personal data use                                                                        &                                                                                                                                                            \\
Passwords                                                                                &                                                                                                                                                            \\
Software updates                                                                         &                                                                                                                                                            \\ \bottomrule

\end{longtable}
\end{landscape}

\end{document}